\documentclass{webofc}
\usepackage[varg]{txfonts}   %
\usepackage[hidelinks]{hyperref}

\usepackage{cleveref}

\crefname{figure}{Fig.}{Figs.}
\Crefname{figure}{Figure}{Figures}
\crefname{equation}{Eq.}{Eqs.}
\Crefname{equation}{Equation}{Equations}
\crefname{section}{Sec.}{Secs.}
\Crefname{section}{Section}{Sections}
\creflabelformat{equation}{#2\textup{#1}#3}

\begin{document}

\title{\bfseries\boldmath High-$p_T$ Suppression in Small Systems}

\author{\firstname{Coleridge} \lastname{Faraday}\inst{1}\fnsep\thanks{\email{frdcol002@myuct.ac.za}} \and
        \firstname{W.\ A.} \lastname{Horowitz}\inst{1}\fnsep\thanks{\email{wa.horowitz@uct.ac.za}}}

\institute{Department of Physics\char`,{} University of Cape Town\char`,{} Private Bag X3\char`,{} Rondebosch 7701\char`,{} South Africa}

\abstract{%
  We present first results for leading hadron suppression in small collision systems, from a convolved radiative and collisional pQCD energy loss model which receives a short path length correction to the radiative energy loss. We find that the short path length correction is exceptionally large for light flavor final states in both small and large collision systems, due to the disproportionate size of the correction for gluons. We examine various assumptions underlying the energy loss model through the calculation of energy loss weighted expectation values of ratios assumed small by the various assumptions. This calculation shows that the large formation time assumption, which is utilized by most contemporary energy loss models, is invalid for a large portion of the phenomenologically relevant parameter space.
}
\maketitle
\section{Introduction}
\label{intro}

The modification of the spectra of high-$p_T$ hadrons in heavy-ion collisions, measured at RHIC \cite{PHENIX:2001hpc, STAR:2003pjh} and later at the LHC \cite{ALICE:2010yje,ATLAS:2022kqu}, are valuable probes of the quark gluon plasma (QGP) which is formed in such collisions. This modification is qualitatively described by perturbative QCD (pQCD) semi-classical energy loss models \cite{Wicks:2005gt, Dainese:2004te, Schenke:2009gb}, which ascribe the suppression to energy loss of high-$p_T$ partons due to interactions with the QGP.

More recently signatures of QGP formation have been observed in small collision systems such as $p+\mathrm{Pb}$, $d+\mathrm{Au}$ and even high multiplicity $p+p$ collisions. These signatures include collective effects  \cite{ALICE:2023ulm, CMS:2015yux}, quarkonium suppression \cite{ALICE:2016sdt}, and strangeness enhancement \cite{ALICE:2015mpp}. Notably, however, the modification of high-$p_T$ hadron spectra in small systems appears to be consistent with unity \cite{ATLAS:2022kqu, ALICE:2019fhe}, which is inconsistent with most energy loss models \cite{Huss:2020whe, Ke:2022gkq}. The measurement of nuclear modification in small systems is however plagued with difficulties, predominantly centrality bias \cite{ALICE:2023csm}. The PHENIX collaboration attempts to circumvent this by calculating a nuclear modification factor normalized by prompt photons \cite{PHENIX:2023dxl}, and finds non-trivial suppression of high-$p_T$ hadrons in $d+\mathrm{Au}$ collisions, emphasizing the importance of centrality bias in small systems, and providing evidence for suppression in small systems.

This uncertainty in the measured nuclear modification factor in small systems necessitates a stronger understanding of the theoretical predictions in these small systems. In this work we present novel predictions from an energy loss model \cite{Faraday:2023mmx, Faraday:2023vbo} based on the Wicks-Horowitz-Djordjevic-Gyulassy (WHDG) energy loss model \cite{Wicks:2005gt}, where the Djordjevic-Gyulassy-Levai-Vitev (DGLV) \cite{Djordjevic:2003zk} radiative energy loss receives a short path length (SPL) correction \cite{Kolbe:2015rvk} which is explicitly neglected in all other contemporary energy loss models. The SPL correction to the radiative energy loss \cite{Kolbe:2015rvk} is exponentially suppressed in large systems, grows faster in $p_T$ than the uncorrected DGLV radiative energy loss, decreases the radiative energy loss which allows for energy gain relative to the vacuum, and is contemporary disproportionately large for gluons as opposed to quarks (compared to the usual $C_A / C_F$ color scaling).

In this work we will show that the SPL correction reveals fundamental inconsistencies in contemporary energy loss models.

\section{Results}
\label{sec:results}

\begin{figure}[htb]
  \centering
  \includegraphics[width=0.45\linewidth]{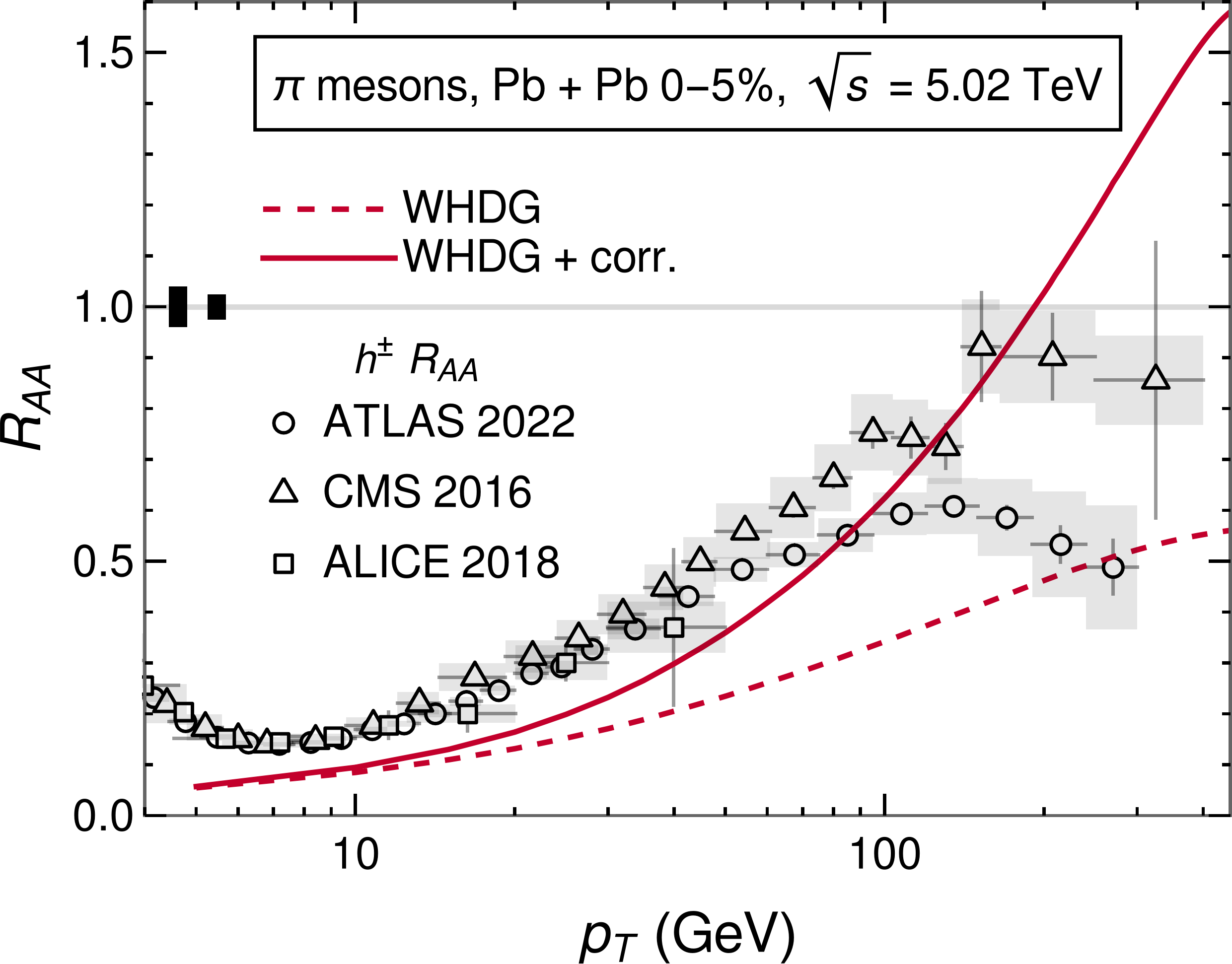}
  \includegraphics[width=0.45\linewidth]{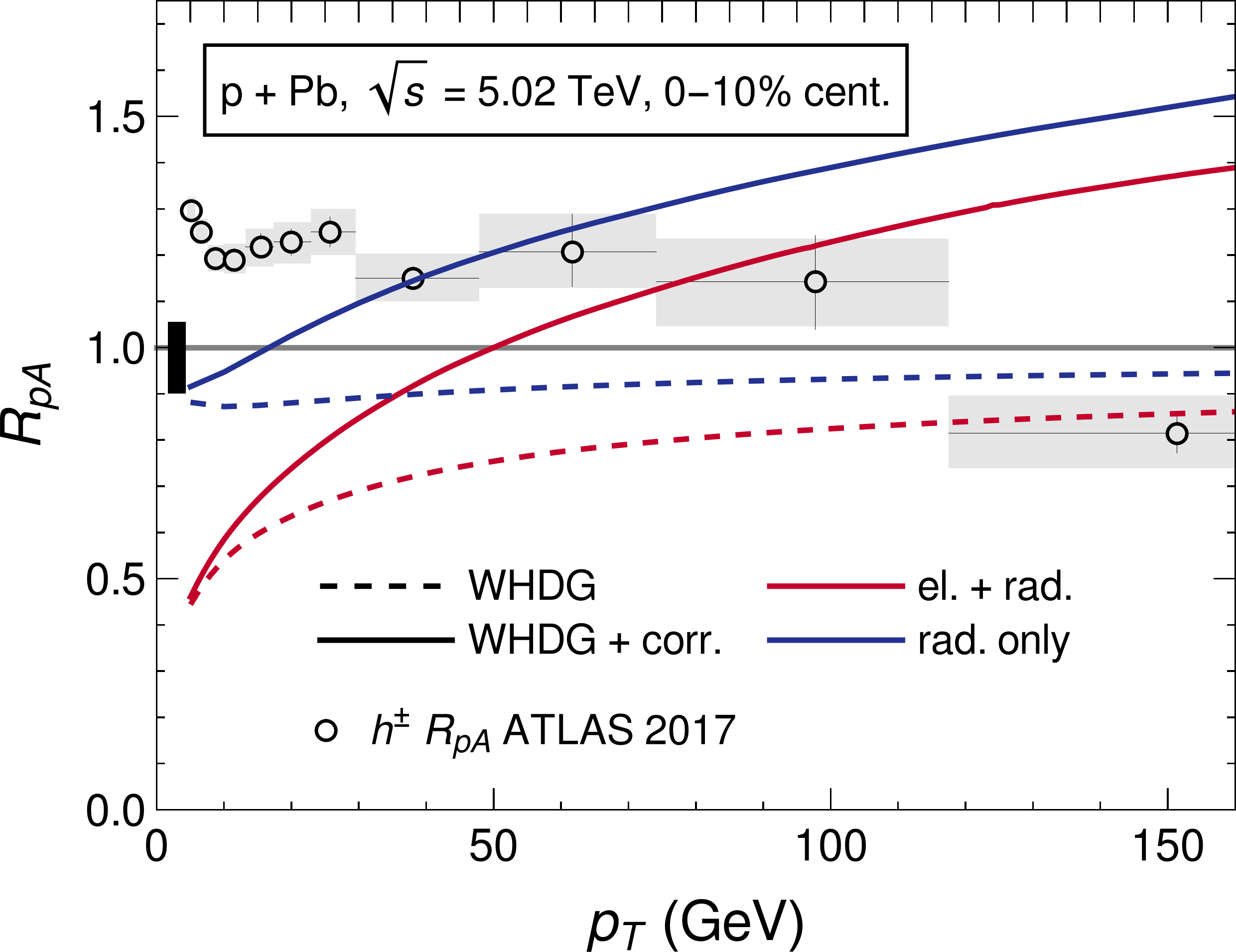}
  \caption{(Left) Nuclear modification factor for pions in $\mathrm{Pb} + \mathrm{Pb}$ collisions with and without the SPL correction, compared with data from ATLAS \cite{ATLAS:2022kqu}, CMS \cite{CMS:2016xef}, and ALICE \cite{Sekihata:2018lwz}. (Right) Nuclear modification factor for pions in $p + \mathrm{Pb}$ collisions with and without the SPL correction, and including and neglecting the collisional energy loss; compared with data from ATLAS \cite{ATLAS:2022kqu}.}
  \label{fig:pion_nuclear_modification}
\end{figure}

In \cref{fig:pion_nuclear_modification} predictions from our energy loss model model \cite{Faraday:2023mmx, Faraday:2023vbo} for the nuclear modification factor of pions is presented for $\mathrm{Pb} + \mathrm{Pb}$ and  $p + \mathrm{Pb}$ collisions with and without the inclusion of the SPL correction. In the left pane of \cref{fig:pion_nuclear_modification}, we note that the SPL correction is large in heavy-ion collisions where we expect it to be small, and that the SPL correction might offer an explanation for the surprisingly fast growth of the nuclear modification factor in $p_T$. In evidence in this figure, is the faster growth of the SPL correction in $p_T$ relative to the uncorrected result, as well as the disproportionate size of the correction for gluons (which largely hadronize into pions). In the right pane of \cref{fig:pion_nuclear_modification}, predictions are made with and without the SPL correction, as well as including and neglecting the collisional energy loss. The figure shows that the collisional energy loss dominates the energy loss in $p+\mathrm{Pb}$ collisions, which implies the need for an collisional short path length correction in order to make quantitative predictions. The SPL corrected results predict enhancement, which is consistent with data for $\mathcal{O} (10\text{--}100)$ GeV, however the growth in $p_T$ past $100$ GeV seems inconsistent with data.

The excessive size of the SPL correction for pions produced in central heavy-ion collisions, where one would expect the correction to be small, leads us to question the assumptions underlying the energy loss model. We systematically check all the assumptions in the model by calculating the energy loss weighted expectation value of dimensionless ratios which are assumed to be much smaller than one in our model. Specifically if it is assumed that $R \ll 1$ in our model, we calculate
$
\langle R \rangle \equiv \int \mathrm{d} \{X_i\} ~ R(\{X_i\}) \; \left | \frac{\mathrm{d} E}{\mathrm{d} \{ X_i \}} \right | {\bigg/} \int \mathrm{d} \{X_i\}~ \left | \frac{\mathrm{d} E}{\mathrm{d} \{X_i\}} \right |
$,
where $dE / d\{X_i\}$ is the radiative energy loss kernel and $\{X_i\}$ are various quantities which the ratio might depend on such as the radiated transverse gluon momentum $\mathbf{k}$, the radiated momentum fraction $x$, and the transverse momentum exchanged with the medium $\mathbf{q}$. We calculate the expectation of ratios assumed small by softness $x \ll  1$, collinearity $k^- / k^+ \ll  1$, and LFT $\mathbf{k}^2 / 2x E \mu \ll  1$ where $E$ is the energy of the hard parton and $\mu$ is the Debye mass. We find that while softness and collinearity are satisfied self-consistently, the LFT assumption is explicitly violated at moderate $\mathcal{O} (10\text{--}100)$ GeV momenta and $\mathcal{O} (1\text{--}5)$ fm lengths.

\begin{figure}[htb]
  \centering
  \includegraphics[width=0.45\linewidth]{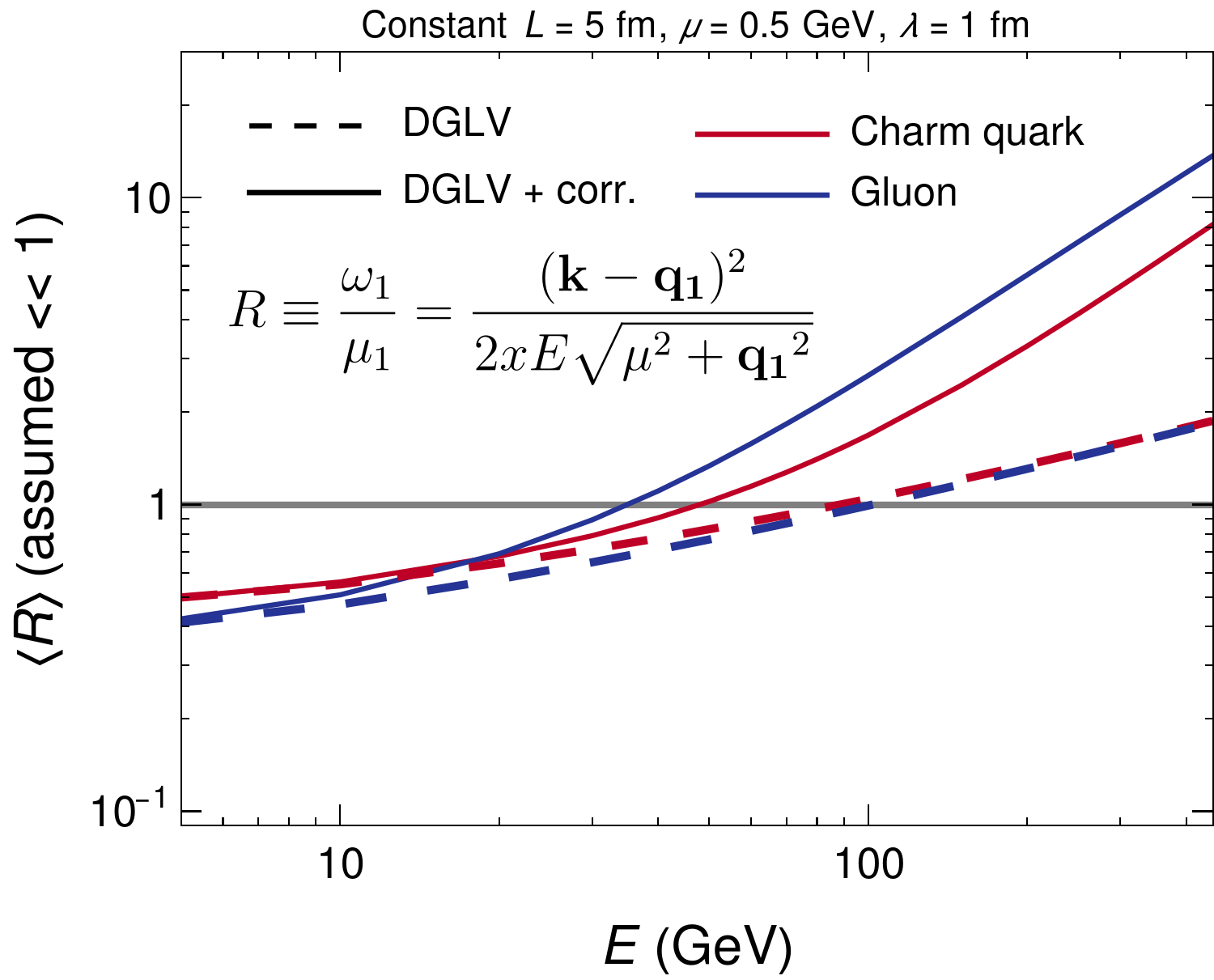}
  \includegraphics[width=0.45\linewidth]{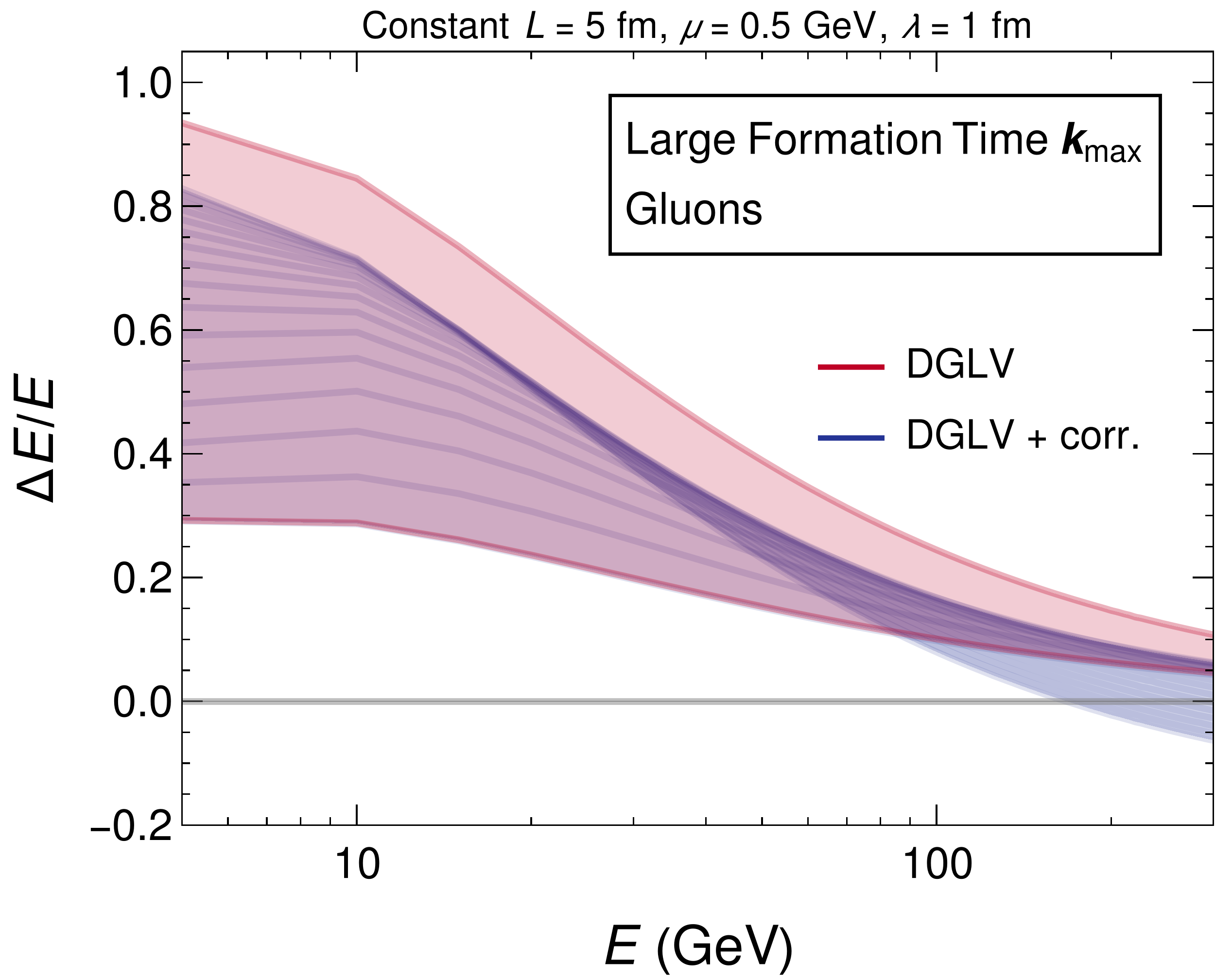}
  \caption{(Left) Energy weighted expectation of a ratio assumed small by the LFT assumption, for the DGLV result and the DGLV result with the SPL correction. (Right) Fractional energy loss for DGLV and DGLV with SPL correction, with a kinematic cut on the radiated gluon transverse momentum determined such that the energy loss kernel is never evaluated in regions where the LFT assumption is invalid. The bands result from varying the kinematic cut by factors of two.}
  \label{fig:assumption}
\end{figure}

In the left pane of \cref{fig:assumption}, we plot the energy weighted expectation of a ratio $R$ which is assumed small according to the LFT assumption. We observe that the LFT assumption breaks down for both the DGLV result and the DGLV result with the SPL correction, however the breakdown occurs earlier in $p_T$ for the SPL corrected result. In addition the breakdown occurs earlier and is more dramatic for gluons as opposed to quarks, indicating that the pion results at high-$p_T$ in \cref{fig:pion_nuclear_modification}, are receiving large contributions from regions of phase space where the LFT assumption is invalid. 

In the right pane of \cref{fig:assumption} we explore a potential way to control the LFT assumption by imposing a kinematic cut on the radiated gluon transverse momentum, which ensures that the energy loss kernel is never evaluated in regions of phase space for which the LFT assumption is invalid \cite{Faraday:2023uay}. The bands in the figure result from varying the kinematic cut by factors of two, and are indicative of the theoretical uncertainty resulting from this prescription. Including this kinematic cut dramatically reduces the size of the SPL correction, however the sensitivity to the precise choice of kinematic cut is large for both the corrected and uncorrected radiative energy loss.

\section{Conclusion}
\label{sec:conclusion}

We have presented first results for leading hadron suppression predictions in both $p + A$ and $A + A$ collisions from a convolved radiative and collisional energy loss model, which receives a short path length correction to the radiative energy loss. We find that the SPL correction is exceptionally large for light flavor final states in both small and large collision systems, due to the disproportionate size of the correction for gluons, prompting a thorough examination of the assumptions underlying the model. We observe that the large formation time assumption, which is utilized by most energy loss models, is violated at moderate $\mathcal{O}(10 \text{--}100)$ GeV momenta and $\mathcal{O}(1 \text{--} 5)$ fm lengths.

Direct future work could include a large formation time correction to the DGLV radiative energy loss, small system size corrections to the collisional energy loss, and predictions resulting from the LFT kinematic cut.
Additional future work in small systems may seek to place energy loss calculations on a more rigorous footing \cite{Clayton:2021uuv} or consider the small system size corrections to thermodynamics \cite{Mogliacci:2018oea}, the equation of state \cite{Horowitz:2021dmr}, and the effective coupling \cite{Horowitz:2022rpp}.

\section*{Acknowledgements}
\label{sec:acknowledgements}

CF and WAH thank the South African National Research Foundation and SA-CERN Collaboration for financial support. CF thanks CERN for hospitality during this work.

\end{document}